\documentclass[letterpaper,12pt]{article}
\usepackage{epsfig}
\usepackage{graphicx}
\usepackage{subfigure}
\newcommand{\be}{\begin{equation}}
\newcommand{\ee}{\end{equation}}
\newcommand{\bea}{\begin{eqnarray}}
\newcommand{\eea}{\end{eqnarray}}
\newcommand{\bml}{\begin{mathletters}}
\newcommand{\eml}{\end{mathletters}}

\def\nonu{\nonumber}
\def\br{\begin{eqnarray}}
\def\er{\end{eqnarray}}
\def\be{\begin{equation}}
\def\ee{\end{equation}}
\def\({\left(}
\def\){\right)}
\def\pa{\partial}
\def\u2{\mid u\mid^2}
\def\rlx{\relax\leavevmode}
\def\IR{\rlx\hbox{\rm I\kern-.18em R}}
\def\IZ{\rlx\hbox{\sf Z\kern-.4em Z}}

\def\vp{\varphi}

\newcounter{fixy}
\begin{document}
\newenvironment{fixy}[1]{\setcounter{figure}{#1}}
{\addtocounter{fixy}{1}}
\renewcommand{\thefixy}{\arabic{fixy}}
\renewcommand{\thefigure}{\thefixy\alph{figure}}
\setcounter{fixy}{1}





\title{Wobbles and other kink-breather solutions of the Sine Gordon model}

\author{L. A. Ferreira\thanks{e-mail address: laf@ifsc.usp.br}
\\ Instituto de F\'isica de S\~ao Carlos, Universidade de S\~ao Paulo\\
Caixa Postal 369, 13.560-970, S\~ao Carlos, SP, Brazil\\
\\ Bernard Piette\thanks{e-mail address:
  B.M.A.G.Piette@durham.ac.uk}\qquad  and \quad  
 Wojtek J. Zakrzewski\thanks{e-mail address: W.J.Zakrzewski@durham.ac.uk} 
\\ Department of Mathematical Sciences, \\ University
of Durham, Durham DH1 3LE, U.K.}
 
\date{\today}
\setlength{\footnotesep}{0.5\footnotesep}

\maketitle
\begin{abstract}
We study various solutions of the Sine Gordon model in (1+1) dimensions.
We use the Hirota method to construct some of them and then show that 
the wobble, discussed in detail by K\"alberman \cite{Kalberman}, is
one of such solutions. 
We concentrate our attention on a kink and its bound states with one or two
breathers. We study their stability and  some aspects of their
scattering properties on potential 
wells and on fixed boundary conditions.

\end{abstract}
\renewcommand{\thefootnote}{\arabic{footnote}}
\section{Introduction}

Topological solitons play an important role in the description of many
phenomena  
in physics. In this paper we look at solitons of the simplest model in (1+1) 
dimensions, namely the Sine-Gordon model.

This model involves a scalar field $\varphi(x,t)$ and is based on the
Lagrangian density 
given by (we set the speed of light to $c=1$) 
\begin{equation}
L\,=\,\frac{1}{2}(\frac{d\varphi}{dt})^2\,-\,\frac{1}{2}
(\frac{d\varphi}{dx})^2\,-\frac{m^2}{\beta^2}
\left[1-\cos({\beta\,\varphi})\right].  
\label{one}
\end{equation}

This particular model arises in many areas which range from the
description of Josephson junctions \cite{Jos} 
to systems with one-dimensional dislocations \cite{dis}. The model has
also been very intensively studied by 
mathematicians (as it describes spaces with constant negative
curvature \cite{geom}) and by theoretical 
physicists working in integrable and conformal field theories \cite{conf}.

As is well known \cite{Abl} the model possesses  kink, antikink  and
breather solutions. A kink solution 
is a static field configuration
which solves the Euler Lagrange equations based on (\ref{one})
{\it ie} 
\begin{equation}
\varphi_{tt}\,-\,\varphi_{xx}\,=\,-\frac{m^2}{\beta} \sin(\beta\varphi)
\label{two}
\end{equation} 
and satisfies the boundary conditions $\varphi(x=-\infty)=0$ and
$\varphi(x=\infty)=2\pi$. 
Such a field is well known \cite{Abl} and is given  by
\begin{equation}
\label{three}
\varphi\,=\,\frac{4}{\beta}\, {\rm ArcTan} \left[\exp(m\,(x-x_0))\right]
\end{equation}
For the antikink the boundary conditions are interchanged and in the
field configuration given above 
there is a `-' sign before ($x-x_0$).

In addition, the model possesses  also so-called `breather' solutions. 
These are nonstatic field  solutions of (\ref{two}) given by
\begin{equation}
\label{four}
\varphi\,=\,\frac{4}{\beta}\, {\rm ArcTan}
\left[\frac{\sqrt{1-\omega^2}}{\omega}\frac{\sin(m\,\omega
    t)}{\cosh(m\,\sqrt{1-\omega^2}(x-x_0))} 
\right].
\end{equation}
Here $\omega$ is a free parameter of the solution which varies from -1
to 1. The breather field, which 
can be thought of as describing a bound state of a kink and an
antikink, oscillates  
with frequency $m\,\omega$.

As the basic Lagrangian is Lorentz covariant - all these field
configurations can be Lorentz boosted 
resulting in field configurations moving with velocity $v<c=1$.

Recently there has been some controversy as to whether a kink
possesses an internal mode 
\cite{aa} - \cite{bb}. 
Such a mode was claimed to exist by Boesch and Willis \cite{Boesch}
and then disclaimed by Quintero et al \cite{Mertens}. The mode at
stake is a possible internal mode of zero frequency which may have
arisen in 
numerical studies of the Sine-Gordon model. Such a mode could be a
genuine oscillatory mode or a numerical 
artifact.

In fact, as is well known \cite{Abl}, the Sine-Gordon model possesses
many solutions in addition to above mentioned  
kinks and breathers. One such solution was recently studied
extensively by G. K\"alberman \cite{Kalberman}. 
He called it `a wobble' and looked at its properties in
detail. However, it is not clear from his 
discussion whose claims his wobble solution supports.

It is worth recalling at this stage that the wobble is only one of 
many solutions which involve a kink with breathers.
Hence we have decided to revisit this subject and look in some detail
at these solutions. 
In particular, we have decided to present explicit forms of the field
configurations which describe some of 
these solutions, so that they become better known outside the
`integrable model community'. 
We have also decided to look at some of their properties, paying particular
attention to their stability 
and their scattering properties on defects (here taken in the form of
an interaction with potential holes 
and boundaries). This we discuss in the next sections.

\section{The wobble}
\label{sec:wobblemain}

The wobble solution of K\"alberman \cite{Kalberman} involves a field
configuration describing a static kink 
and a breather. In his paper K\"alberman gives an analytic form of this
solution and then discusses some of its 
properties. The analytical form of this solution was obtained  by
K\"alberman using the Inverse scattering method of Lamb and Segur.  

In our work we use the Hirota method \cite{hirota} of deriving such
solutions as discussed in detail in the
appendix. We derive in section \ref{sec:wobbleapp} the exact solution
describing a kink and a breather moving with respect to each other. The
solution is given by eq. (\ref{kink+breather}). However, the wobble
corresponds to a kink and a breather sitting on top of each other 
and not having a relative motion. Therefore, if one sets the the
velocities to zero in the general solution  (\ref{kink+breather}) one
gets the wobble solution as 
\be
\vp = \frac{4}{\beta}\; {\rm ArcTan}\;\frac{\left[
    \frac{\sqrt{1-\omega^2}}{\omega}\,\sin \(m\, \omega\, t\)  +   
\frac{1}{2}\, e^{\varepsilon\, m\, x}\( e^{-m\,\sqrt{1-\omega^2}\, x} + 
 \rho^2\; e^{m\,\sqrt{1-\omega^2}\,x}\)\right]}{\left[  \cosh\(
    m\, \sqrt{1-\omega^2}\,x\) 
+ \frac{\sqrt{1-\omega^2}}{\omega}\,\rho\, e^{ \varepsilon\, m\, x} \,
\sin\(m\,\omega\, t\)\right]} 
\label{staticbreather}
\ee
where $\omega$ is a frequency varying from $-1$ to $1$, and 
\be
\rho =\frac{1-\varepsilon\, \sqrt{1-\omega^2}}{1+\varepsilon\,
  \sqrt{1-\omega^2}}  
\label{defs}
\ee
where $\varepsilon=\pm 1$ corresponds to the kink ($+1$) or anti-kink
($-1$). 

This agrees with the expression given by K\"alberman. 
As is clear from (\ref{staticbreather}) the field configuration
depends on one parameter (the frequency 
of the breather) and so we have studied the stability of this field
configuration  
by calculating $\vp$ and its time derivative from
(\ref{staticbreather}) and then used the fourth-order 
Runge Kutta method to simulate the time evolution of this
configuration (the spatial derivatives 
were calculated using central differences). 
Our simulations involved looking at a breather-kink system ($\varepsilon=1$)
and for $\beta$ and $m$ we took $\beta=m=1$. The results of our
simulations were in complete 
agreement with the analytical expression thus showing that the
solution is stable with respect 
to small perturbations (due to the discretisations). 

Next we tried to assess the stability of the wobble with respect to
larger perturbations. 
We have performed several perturbations, the most important of them being 
the perturbation of the original slope of the kink. ({\it ie} in the
expression (\ref{staticbreather})  
we have replaced $\exp(\varepsilon m x)$ by $exp(\lambda x)$ where
$\lambda\ne1$). We 
have performed numerical simulations 
with $\lambda=1.05,1.15, 1.18, 1.20,1.24$ and $1.3$.

Each perturbation added an extra energy to the system. Such a system
was then unstable and so it evolved towards a stable wobble emitting
some radiation which was sent out towards the boundaries of the
grid. To prevent the reflections 
from the boundaries we absorbed the energy there.  For $\lambda$ close
to one - the perturbations were small - hence the system returned to
its initial configuration 
(with $\lambda=1.0$). For larger values of the perturbation the system
was more perturbed and often  
not only kept on sending out its excess of energy but also,
at regular intervals, altered its frequency of oscillation (increasing
it) which allowed it to 
send out even more radiation.
In fig 1. we present the plots of the time dependence of the total
energy 
 as seen in the simulation in which $\lambda$ was set at 1.15.

\begin{fixy}{-1}
\begin{figure}
\centering
\epsfysize=10cm
\mbox{\epsffile{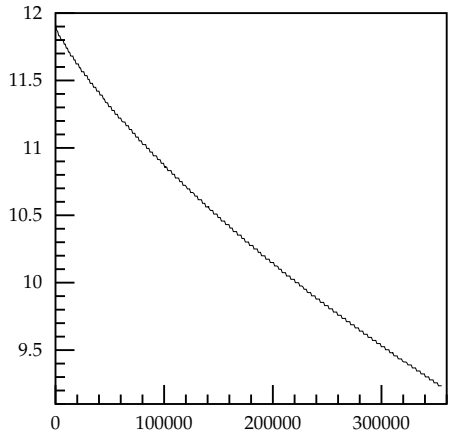}}
\label{fig1}
\caption{\label{Fig.1} Energy as a function of time as seen in a
  simulation started with  
$\lambda=1.15$}
\end{figure}
\end{fixy}

We have also studied the case of $v=v_k=v_B\ne0$. In this case we had
a system consisting 
of a kink and a breather moving with a constant velocity. Again, the
system was stable 
(we run it with very  small $v$ to avoid having problems with the boundaries
and for larger values of $v$ with fixed boundary conditions).

Then we performed a series of simulations in which the initial
configuration was sent towards a potential hole. This was achieved by
making $\alpha\equiv \frac{m^2}{\beta^2}$ in (\ref{one}) $x$
dependent; {\it i.e.} we set  
\begin{equation}
\alpha(x)=1,\quad \hbox{for}\quad \vert x\vert >5,\quad
\alpha(x)=\alpha_0<1\quad \hbox{for} \quad -5<x<5.\label{nine} 
\end{equation}
Then placing the breather and the kink long way away from the hole
({\it ie} from $-5<x<5$), 
and sending them towards it we could study the effects of their
scattering on the hole. 
 We have found that the hole can separate the breather from the kink
 (in one simulation 
we saw the kink being trapped in the hole while the breather bounced
off the kink  
trapped in the hole and returned to where the system has originally come from).
As studied by two of us \cite{Bernard}  the scattering of a breather
on the hole 
is very complicated and produces many different outcomes; this time we
have even more  
possibilities and so  we have decided to postpone the further study of this
problem to some future work.

\section{Kink and two breathers}
\label{sec:kink+twobreathers}

In this section we briefly discuss another interesting solution of the
sine-Gordon model; namely 
the solution corresponding to one static  kink and two breathers.

In this case, as shown in the appendix, the field is given, for $\beta=m=1$, by
\begin{equation}
\vp = 4\; {\rm ArcTan}\;\frac{{\cal A}}{{\cal B}}
\label{twobreathers}
\end{equation}
where
\br
{\cal A}&=& -2\, {\rm cotan}\, \theta_1\, e^{x\; \cos \theta_1}\,
\sin\(t\; \sin \theta_1\)  
 -2\, {\rm cotan}\, \theta_2\, e^{x\; \cos \theta_2}\, \sin\(t\; \sin
 \theta_2\)   
+ e^{x }
\nonu\\
&-& 2\, {\rm cotan}\, \theta_2\; \sigma_{12}^{(+)}\, \sigma_{12}^{(-)}\,
e^{x\; \(2\,\cos \theta_1+\cos \theta_2\)}\, \sin\(t\; \sin \theta_2\) 
+ \rho_1^2   \; e^{ 2\, x\; \cos \theta_1} \;  e^{x } \nonu\\
&-&  2\, {\rm cotan}\, \theta_1\; \sigma_{12}^{(+)}\, \sigma_{12}^{(-)}\,
e^{x\; \(\cos \theta_1+2\,\cos \theta_2\)}\, \sin\(t\; \sin \theta_1\)   
+ \rho_2^2   \; e^{ 2\, x\; \cos \theta_2} \;  e^{x }    \nonu\\
&+&  2\, \sigma_{12}^{(-)}\, \rho_1\, \rho_2\, {\rm cotan}\,
\theta_1\,{\rm cotan}\, \theta_2 
\, e^{\varepsilon\; x}\, e^{x\; \(\cos \theta_1+\cos \theta_2\)}\, 
\cos\(t\, \(\sin \theta_1+\sin \theta_2\)\) \nonu\\
&-&  2\, \sigma_{12}^{(+)}\, \rho_1\, \rho_2\, {\rm cotan}\,
\theta_1\,{\rm cotan}\, \theta_2 
\, e^{\varepsilon\; x}\, e^{x\; \(\cos \theta_1+\cos \theta_2\)}\, 
\cos\(t\, \(\sin \theta_1-\sin \theta_2\)\)
\nonu\\
&+&\(\sigma_{12}^{(+)}\,\sigma_{12}^{(-)}\,\rho_1\, \rho_2\)^2\; e^{x }\, 
e^{2\,x\; \( \cos \theta_1+\cos \theta_2\)}
\er
and
\br
{\cal B}&=& 1+ e^{2\,x\; \cos \theta_1} + e^{2\,x\; \cos \theta_2}
+ 2\, \rho_1\, {\rm cotan}\, \theta_1\, e^{x }\, e^{x\; \cos
  \theta_1}\, \sin\(t\; \sin \theta_1\)  
\nonu\\
&+&  2\, \rho_2\, {\rm cotan}\, \theta_2\, e^{x }\, e^{x\; \cos
  \theta_2}\, \sin\(t\; \sin \theta_2\)  
\nonu\\
&+& 2\, \sigma_{12}^{(-)}\, {\rm cotan}\, \theta_1\,{\rm cotan}\,
\theta_2\,e^{x\; \(\cos \theta_1+\cos \theta_2\)}\,  
\cos\(t\, \(\sin \theta_1+\sin \theta_2\)\)
\nonu\\
&-& 2\, \sigma_{12}^{(+)}\, {\rm cotan}\, \theta_1\,{\rm cotan}\,
\theta_2\,e^{x\; \(\cos \theta_1+\cos \theta_2\)}\,  
\cos\(t\, \(\sin \theta_1-\sin \theta_2\)\)
\nonu\\
&+& \(\sigma_{12}^{(+)}\,\sigma_{12}^{(-)}\)^2 \, e^{2\;x\; \(\cos
  \theta_1+\cos \theta_2\)} \nonu\\ 
&+&  2\, \sigma_{12}^{(+)}\,\sigma_{12}^{(-)}\, \rho_1^2\, \rho_2\,
    {\rm cotan}\, \theta_2\, e^{x }\,  
e^{x\; \(2 \cos \theta_1+\cos \theta_2\)}\, \sin\(t\; \sin \theta_2\) \nonu\\
&+&  2\, \sigma_{12}^{(+)}\,\sigma_{12}^{(-)}\, \rho_1\, \rho_2^2\,
{\rm cotan}\, \theta_1\, e^{x }\,  
e^{x\; \( \cos \theta_1+2\,\cos \theta_2\)}\, \sin\(t\; \sin \theta_1\).
\er

Here 
\begin{equation}
\rho_i\,=\, - \;\frac{1-\, \cos \theta_i}{1+\, \cos \theta_i},
\quad \sigma_{12}^{(\pm)}\,=\, -\; \frac{1- \cos
  \(\theta_1\pm\theta_2\)}{1+ \cos\(\theta_1\pm \theta_2\)} 
\end{equation}

This solution depends on two constants $(\theta_1$, $\theta_2$), which
control the frequencies of breathers' 
oscillations.

We have tested the stability of this solution by using the expression
(\ref{twobreathers}) to calculate  
$\varphi(t=0)$ and $\frac{d\varphi}{dt}(t=0)$ and then performing a
numerical simulation of (\ref{two}). 
As before, the discretisation has produced a small perturbation but
the field configuration was stable; 
{\it ie} after a long simulation (we run it till $t=5000$) the field
was indistinguishable from the expression 
(\ref{twobreathers}) and there was no noticeable radiation. Hence we
can conclude that this field configuration 
is also stable.

\section{Perturbed Field Configurations}
\label{sec:perturbedkink}
Given that we have many field configurations which resemble perturbed
kinks ({\it ie} which are given 
by kinks and breathers) we have tried to see what happens when one
perturbs a kink and lets it evolve in time. 
We have looked at various perturbations, paying particular attention
to configurations which involved adding to a kink an extra
perturbation of the form 
\be
\label{pert}
\delta\varphi(t=0)\,=\,\frac{B}{\cosh(\mu x)},\quad
\delta\frac{\partial\varphi}{\partial t}(t=0)\,=\,\frac{A}{\cosh(\nu
  x)}. 
\ee
We have looked at various values of $A$, $B$, $\mu$ and $\nu$. In all
cases the perturbation made the kink 
move and generated many moving breather-like configurations. We let
the system evolve - absorbing 
the energy at the boundaries of our grid. This had the effect of
slowing down the kink. 
In fig 2 we present the plots of the total energy, and of the
potential energy, of one such simulation 
(corresponding to the values $A=0.5$, $B=0.5$, $\mu=1.0$ and $\nu=0.2$).
The curve in fig 2a shows a steady decrease in total energy down to close to the value of the energy of one 
stationary kink. In fig 3a we show the blowup of the total energy at the latter times (at values 
at which the curve in fig2a may appear to be constant).
The figures 2b and 3b show the corresponding changes of the potential energy.

\begin{fixy}{-1}
\begin{figure}
\unitlength1cm \hfil
\begin{picture}(16,8)
 \epsfxsize=8cm \put(0,0){\epsffile{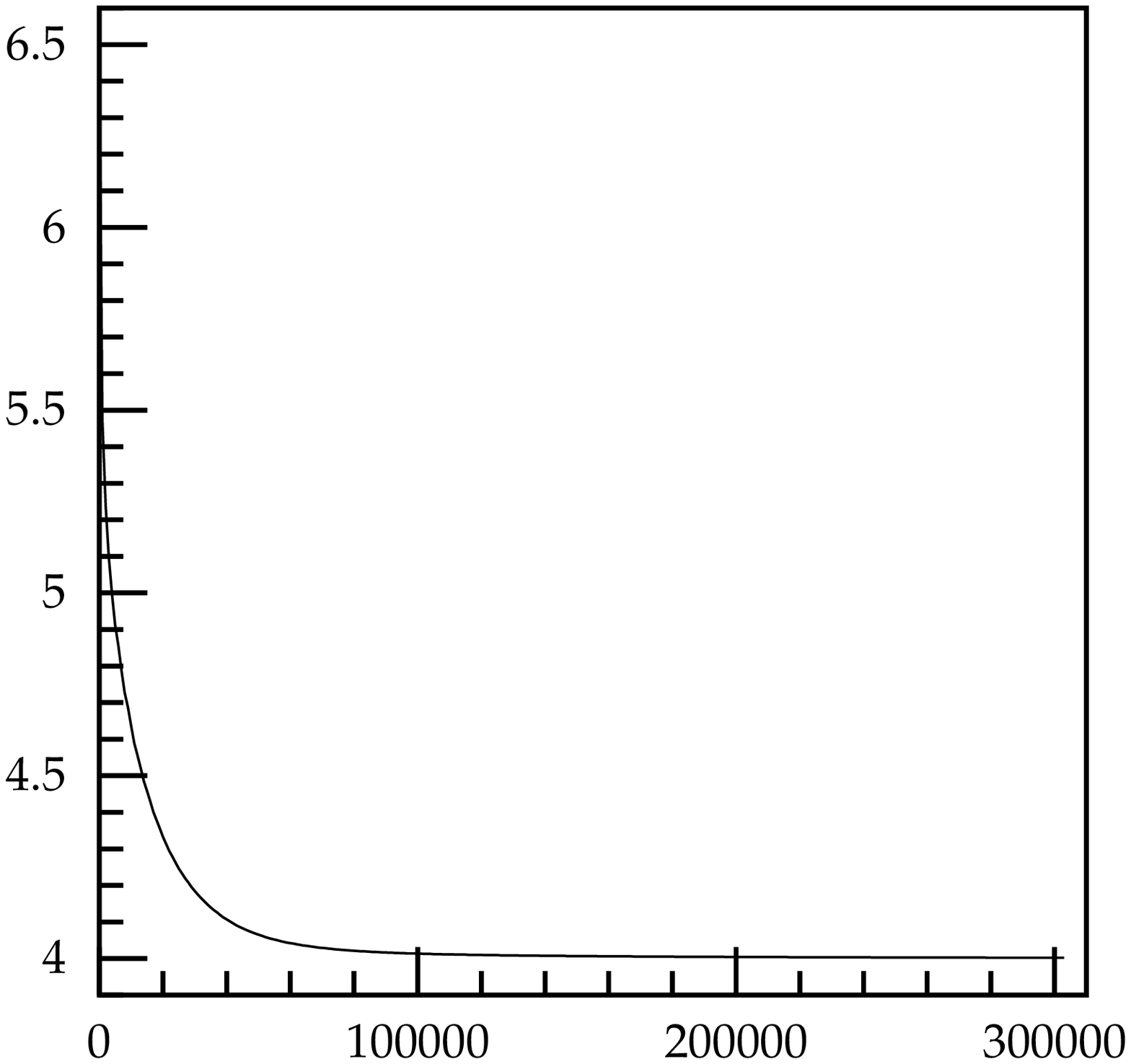}}
 \epsfxsize=8cm \put(8,0){\epsffile{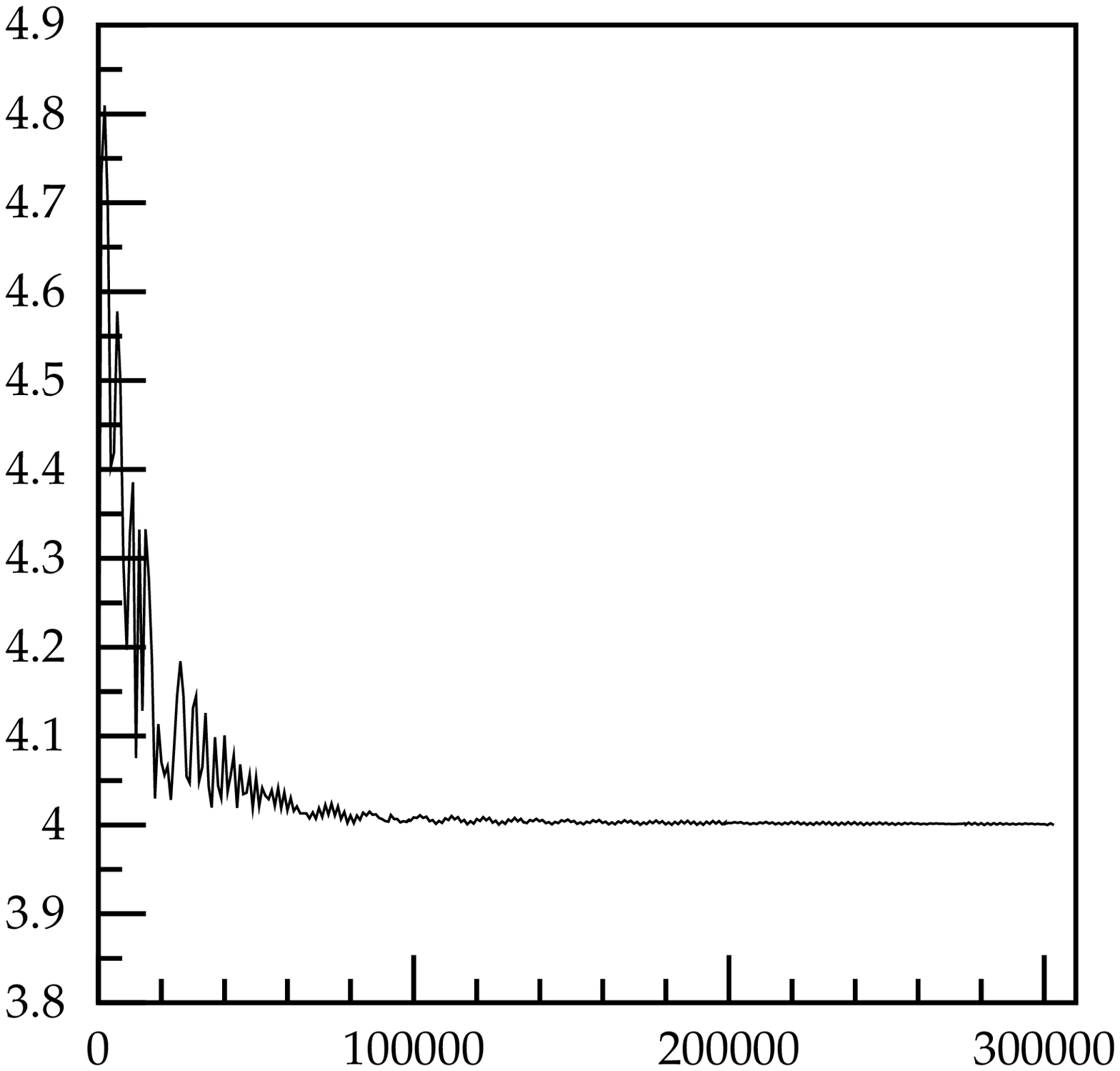}}
\put(4,0){a}
\put(12,0){b}
\end{picture}
\caption{\label{Fig.2} Energies as a function of time as seen when the
  starting file was given by 
(\ref{pert}) 
a) total energy,  
b) potential energy.
}
\end{figure}
\end{fixy}

\begin{fixy}{-1}
\begin{figure}
\unitlength1cm \hfil
\begin{picture}(16,8)
 \epsfxsize=8cm \put(0,0){\epsffile{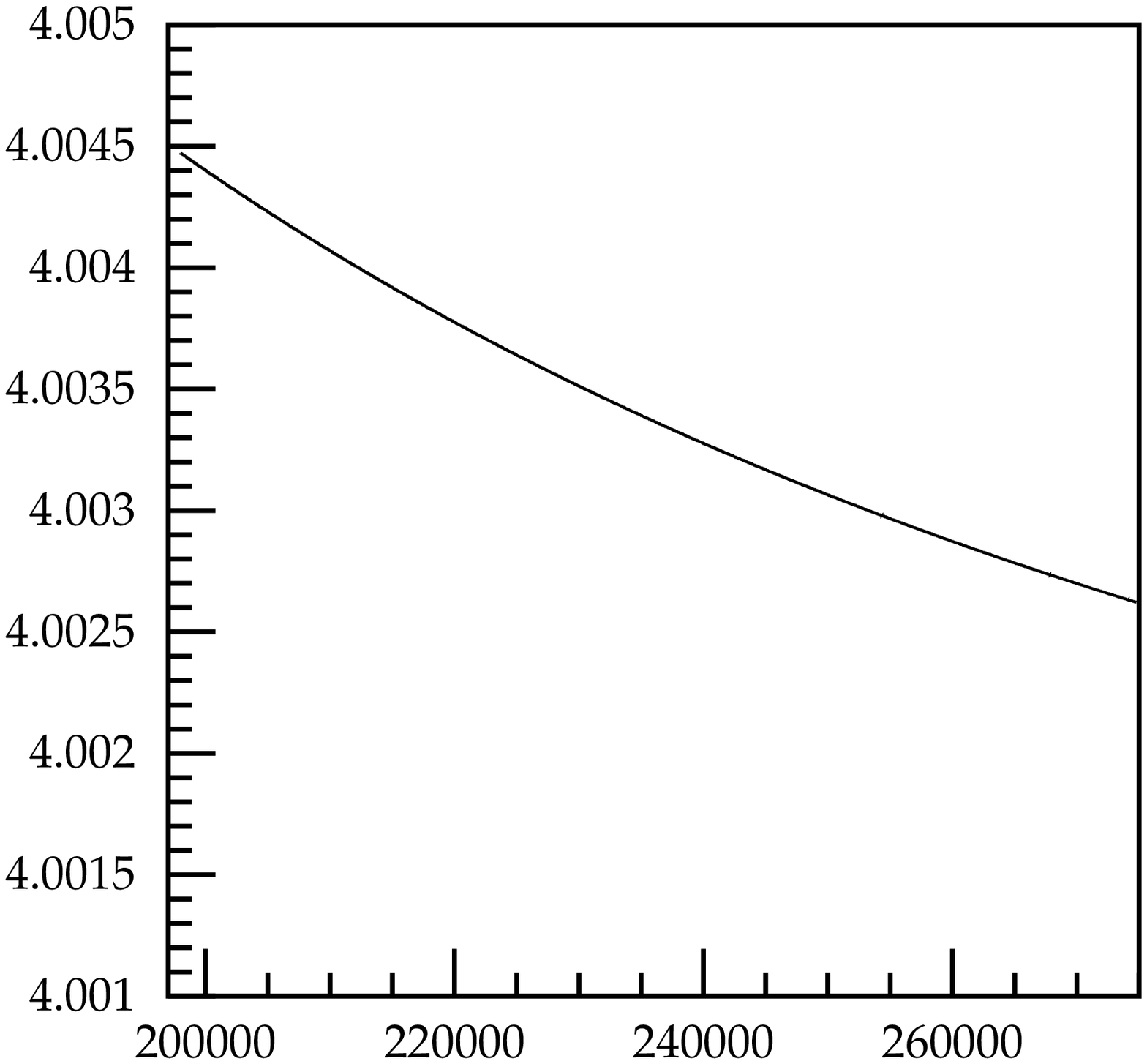}}
 \epsfxsize=8cm \put(8,0){\epsffile{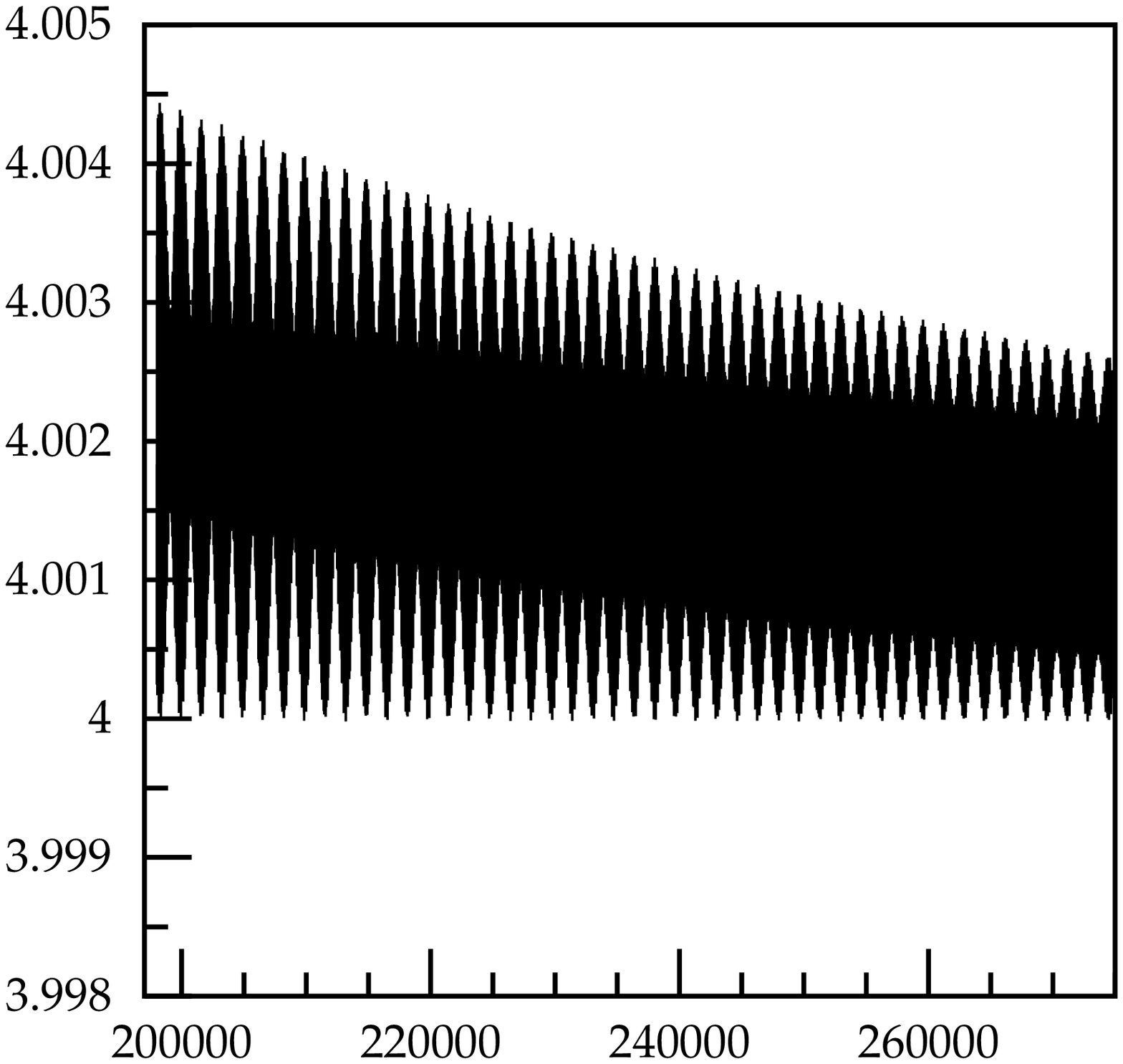}}
\put(4,0){a}
\put(12,0){b}
\end{picture}
\caption{\label{Fig.3} Blowup of the curves of Fig 2 - to show that
  even at later 
times the energies are decreasing.
a) total energy,  
b) potential energy.
}
\end{figure}
\end{fixy}

We note some steps of the decrease
of the total energy (they correspond to the moments when the kink was
reflected from the boundaries). 
The potential has also gradually settled as seen from the plot. Its
oscillation demonstrates the existence 
of transient time dependent structures ({\it ie} breathers). This can
be seen from looking at the time 
dependence of individual field configurations. In fig 4. we present
the plots of the fields at t=6750 and t=6753. 
They show many breather-like structures - the clearest ones being
close to x=35 and x=-38.

\begin{fixy}{-1}
\begin{figure}
\unitlength1cm \hfil
\begin{picture}(16,8)
 \epsfxsize=8cm \put(0,0){\epsffile{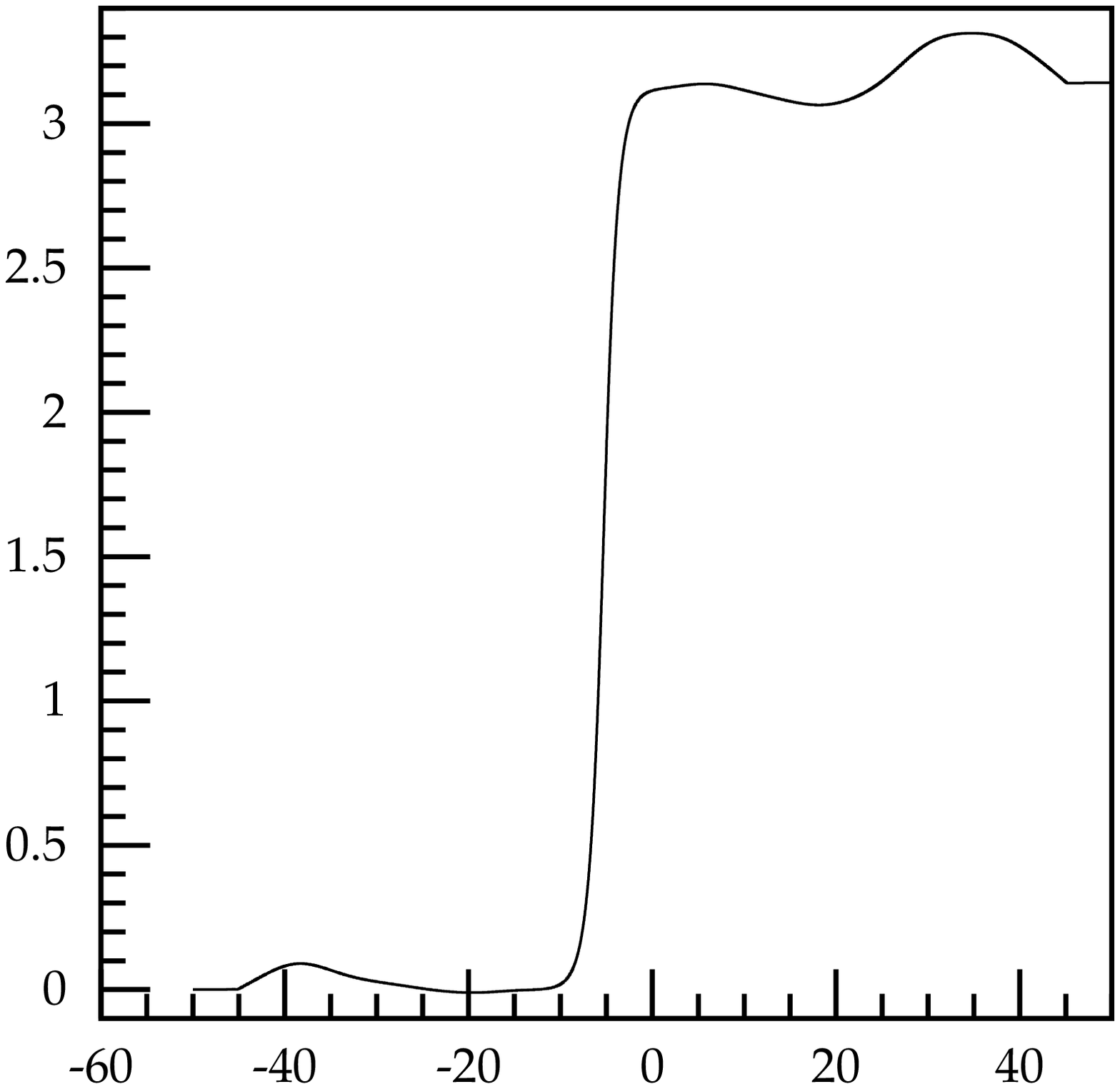}}
 \epsfxsize=8cm \put(8,0){\epsffile{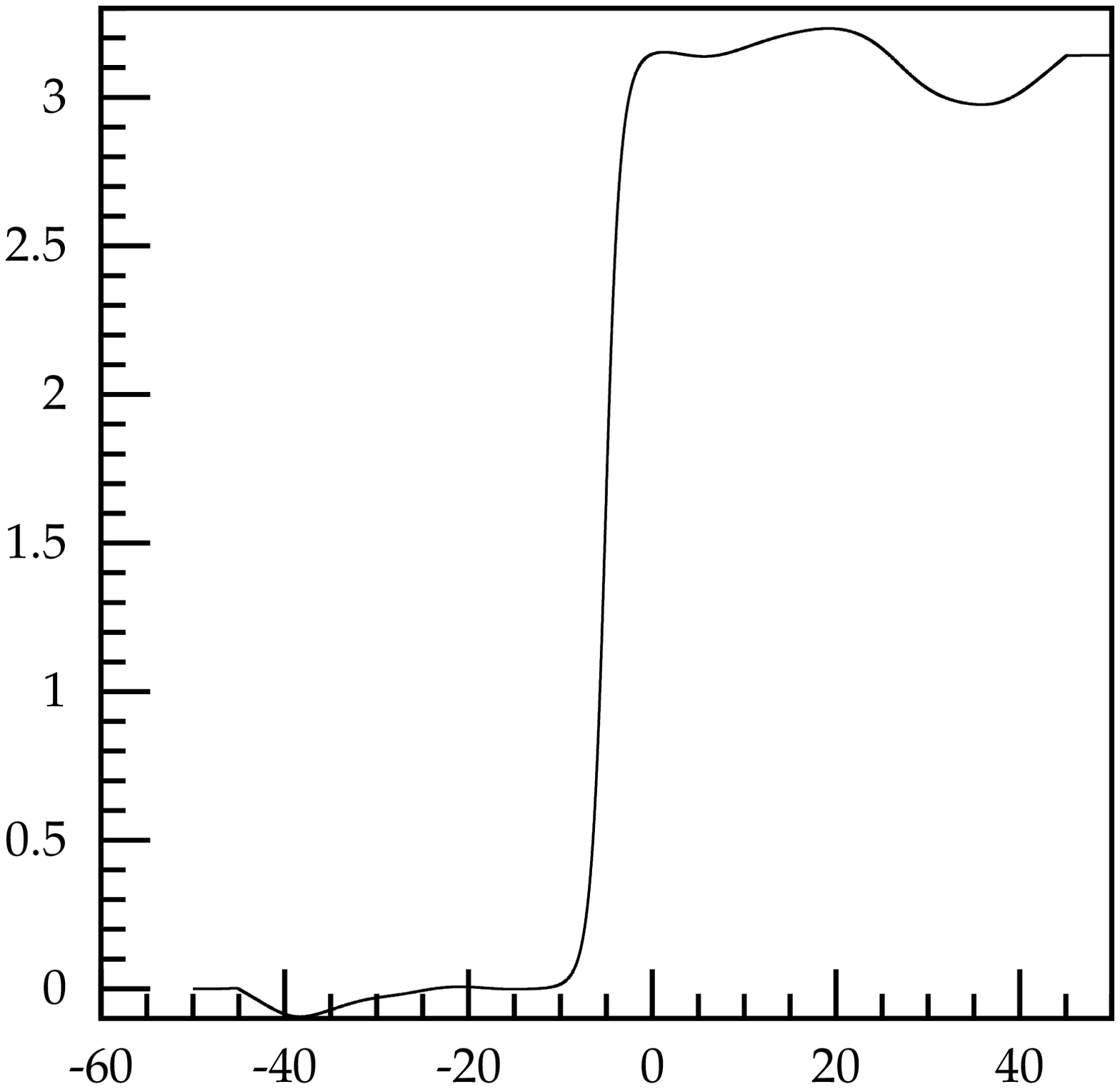}}
\put(4,0){a}
\put(12,0){b}
\end{picture}
\caption{\label{Fig.4}  Field configurations at
 a) t=6750, b) t=6753.
}
\end{figure}
\end{fixy}


Thus it is clear to us that a general field configuration will
gradually split into moving kinks and breathers 
and some radiation which will quickly move out to the
boundaries. However, the resultant field  
configuration is metastable; it still radiates, albeit very slowly,
and gradually evolves towards 
a field configuration involving mainly a kink. Whether at the end of
its evolution we end up with a kink or a kink with some breathers is
hard to determine.

\section{The energy}

In this section we give a simple formula for the energy of the exact
solutions considered in this paper.  
The energy for all the solutions coming from the Hirota method is easy
to calculate since, as we show below  it comes from surface terms. In
addition, the energy is additive for the non-linear Hirota's superposition of
solutions. 
The Hamiltonian density for the sine-Gordon theory (\ref{one}) is
given by 
\be
H\,=\,\frac{1}{2}(\partial_t\varphi)^2\,+\,\frac{1}{2}(\partial_x\varphi)^2\,+\,\frac{2m^2}{\beta^2}\left[\sin\left(\frac{1}{2}\beta\varphi\right)\right]^2,         
\label{ham}  
\ee
Replacing the field $\varphi$ in terms of the Hirota's
$\tau$-functions as given in the appendix (\ref{hirotatau})  we get
\be
H\,=\,-\frac{1}{\beta^2}\left[\left(\frac{\partial_+\tau_1}{\tau_1}-\frac{\partial_+\tau_0}{\tau_0}\right)^2    
\,+\,
\left(\frac{\partial_-\tau_1}{\tau_1}-\frac{\partial_-\tau_0}{\tau_0}\right)^2\,+\, 
\frac{m^2}{2}\left(\frac{\tau_0}{\tau_1}-\frac{\tau_1}{\tau_0}\right)^2\right].
\label{hamil}
\ee
However, this expression can be rewritten as 
\be
H\,=\, \frac{1}{\beta^2}\left(\partial_+-\partial_-\right)^2\left(\ln \tau_0\,+\,\ln \tau_1\right)\,+\, H_{cor}
\label{energy}
\ee
where
\be 
H_{cor}\,=\,
-\frac{1}{\beta^2}\left(\frac{\partial_+^2\tau_0}{\tau_0}\,+\,\frac{\partial_-^2\tau_0}{\tau_0}\,+\,  
\frac{\partial_+^2\tau_1}{\tau_1}\,+\, \frac{\partial_-^2\tau_1}{\tau_1}
\,-\,2\frac{\partial_+\tau_1}{\tau_1}\frac{\partial_+\tau_0}{\tau_0}\,-\,2\frac{\partial_1\tau_1}{\tau_1}\frac{\partial_-\tau_0}{\tau_0}\right). 
\label{correction}
\ee
The  solutions obtained by the Hirota's ansatz (\ref{hirotaansatz}) satisfy,
besides the Hirota's equations (\ref{hirotaeq}), also the additional equations 
\be
\tau_1\partial^2_{\pm}\tau_0 + \tau_0\partial^2_{\pm}\tau_1 - 2\,
\partial_{\pm}\tau_0\, \partial_{\pm}\tau_1 = 0
\label{nicerels}
\ee

This fact is proved, in a much more general setting, in
ref. \cite{wojtek-luiz}. In fact, it is shown there that all
solutions that can be obtained from
the vacuum solution $\varphi=0$ by the dressing
transformations, satisfy (\ref{nicerels}). The Hirota's solutions are,
of course, of this type. Therefore, it turns out that for these
solutions one has $H_{cor}\,=\,0$,
and so
\be 
H\,=\, \frac{4}{\beta^2}\;\partial^2_x\left(\ln \tau_0\,+\,\ln \tau_1\right)
\ee
where we have used the fact that
$(\partial_+-\partial_-)=2\,\partial_x$ (see (\ref{lightcone})).
Therefore, the energy becomes
\be
E=\int_{-\infty}^{\infty} dx\, H = 
\frac{4}{\beta^2}\;\partial_x\left(\ln \tau_0\,+\,\ln
\tau_1\right)\mid_{x=-\infty}^{x=\infty} 
\label{resu}
\ee
and so is determined entirely by the asymptotic values of $\tau_i$ functions.

 In fact, this result, (in a different and less explicit
form) has been known for a while 
for kinks and antikinks,
 to people working in integrable and conformal field theories
 \cite{Olive,Luizandco}. Here we 
 have presented it 
in a form that is more explicit and more easily accessible to people
working in other areas 
of physics. For more details of the general proof see \cite{wojtek-luiz}.
 
Looking at the solutions in the $N$-soliton sector, given in
(\ref{nsolitonsolution}), one observes that the asymptotic behavior of
the $\tau$-function is determined by the exponentials of the
$\Gamma_i$s given in (\ref{gammadef}), {\it i.e.} 
\be
\Gamma_i = \frac{m}{2} \left[\(z_i+\frac{1}{z_i}\)\, x +
  \(z_i-\frac{1}{z_i}\)\, t\right] 
\ee
In (\ref{resu}) we have to evaluate the quantities
$\frac{\partial_x\tau_{\alpha}}{\tau_{\alpha}}\mid_{x=-\infty}^{x=\infty}$.
Therefore, if a given combination of exponentials of the $\Gamma_i$s
dominates the numerator of
$\frac{\partial_x\tau_{\alpha}}{\tau_{\alpha}}$, for $x\rightarrow \pm
\infty$, then the same combination dominates the
denominator. Consequently, when the limit $x\rightarrow \pm \infty$ is
taken, we are left with the ratio of these two dominant terms which
are equal except for the constant term (in the numerator) coming from
the $x$-derivative of the exponentials, {\it i.e.} the terms of the form
$\frac{m}{2}\,\(z_i+\frac{1}{z_i}\)$.  Therefore, the contribution to
(\ref{resu}) is just the terms from the $x$-derivatives. 
In addition,  $\tau_0$
and $\tau_1$, given in (\ref{nsolitonsolution}), have the same form except
for the relative minus signs, and so  their
contributions to (\ref{resu}) are equal, despite these minus signs
which cancel when the limit $x\rightarrow \pm \infty$ is
taken.  

Notice that the solution (\ref{nsolitonsolution}) contains all
possible combinations of the exponentials of the $\Gamma_i$s, such
that each $\Gamma_i$ appears at most once. Therefore, it is clear that
the dominant exponential in the limit $x\rightarrow +\infty$ is
$\exp\(\sum_i \Gamma_i\)$, where in the sum we include all  $\Gamma_i$s
such that ${\rm Re}\(z_i+\frac{1}{z_i}\)>0$. Similarly, the dominant
exponential in the limit $x\rightarrow -\infty$ is 
$\exp\(\sum_i \Gamma_i\)$, where the sum involves all $\Gamma_i$s
such that ${\rm Re}\(z_i+\frac{1}{z_i}\)<0$. Thus, the energy
 depends only on the modulus of ${\rm Re}\(z_i+\frac{1}{z_i}\)$. 
The parameters $z_i$ can be
complex for some solutions and so, writing them as $z_i\equiv
e^{-\alpha_i+i\theta_i}$, one gets
\be
z_i+\frac{1}{z_i}=2
\left[\cos\theta_i\,\cosh\alpha_i-i\,\sin\theta_i\,\sinh\alpha_i\right]. 
\ee
Thus we conclude that the energy (\ref{resu}) for the solutions
(\ref{nsolitonsolution}), is given by 
\be
E = \frac{8\, m}{\beta^2}\; \sum_{i=1}^N 
\left[\; \mid
  \cos\theta_i\mid\,\cosh\alpha_i-i\,\sin\theta_i\,\sinh\alpha_i\right],  
\ee
and where $N$ corresponds to the $N$-soliton sector in which the solution lies. The
energy will then be real for some special choices of the parameters
$z_i$. But whenever this happens, the energy is automaticaly positive. 
Of course, the energy is real whenever the solution $\varphi$ is
real. 

The energies for the solutions we consider in this paper are then given by:
\begin{enumerate}
\item For the $1$-soliton constructed in section \ref{sec:onesol} 
\be
E_{1{\rm -soliton}} =\frac{8\, m}{\beta^2}\frac{1}{\sqrt{1-v^2}}.
\ee
\item For the breather constructed in section \ref{sec:breather} 
\be
E_{{\rm breather}} =\frac{16\,
  m}{\beta^2}\frac{\sqrt{1-\omega^2}}{\sqrt{1-v^2}}. 
\ee
\item For the wobble constructed in sections \ref{sec:wobblemain} and
  \ref{sec:wobbleapp} 
\be
E_{{\rm wobble}} =\frac{8\, m}{\beta^2}\frac{1}{\sqrt{1-v_K^2}}+
\frac{16\,
  m}{\beta^2}\frac{\sqrt{1-\omega^2}}{\sqrt{1-v_B^2}}.  
\ee
\item For the solution of the kink with two breathers  constructed in sections
  \ref{sec:kink+twobreathers} and \ref{sec:kinktwobreathers} (with their
  velocities set to zero)
\be
E_{{\rm kink+2 breathers}} =\frac{8\, m}{\beta^2}+
\frac{16\,
  m}{\beta^2}\sqrt{1-\omega_1^2} +
\frac{16\,
  m}{\beta^2}\sqrt{1-\omega_2^2}, 
\ee
where $\omega_i=\sin \theta_i$, $i=1,2$. 
\end{enumerate}

\section{Final Remarks}
In this paper we have drawn the attention of the readers to the rich structure of solutions (of finite energy)
of the Sine-Gordon model. Even in the one kink sector there are solutions involving, in addition to the kink,
also many breathers. As the energy of each breather depends on its frequency (and vanishes in the limit
of this frequency going to 1) the extra energy, due to these extra breathers, does not have to be very large.
The solutions appear to be stable and this stability is guaranteed by the integrability of the model.
We have tested this numerically and have found that small perturbations, due to the discretisations, do not
alter this stability. To change it we need something more drastic - like the absorption or the space variation 
of the potential ({\it ie} the coefficient of the $sin$ term in the Lagrangian). But even then 
the effects are not very large - one sees splitting of breathers {\it etc} but no `global annihilation'.

At the same time we have looked at the total energy and the energy density of a general solution.
We point out that the total energy is determined by the asymptotic values of the fields
({\it ie} is given by (\ref{resu})). This result has been known earlier (for kinks and antikinks)
to people working in integrable and conformal field theories but may
not be known generally. We have given a general proof of this in a
separate, rather technical, paper \cite{wojtek-luiz}   
but we have also checked it explicitly for all field configurations
involving up to 5 kinks (or antikinks).  

Our results do not answer, definitively, the question as to whether a
kink possesses an internal mode 
of oscillation or not. The perturbation of the kink we performed in
section \ref{sec:perturbedkink} did not produce any oscillatory
internal mode. On the contrary, all the energy given to the kink by the
perturbation was used to produce breather like excitations which died
away very slowly. In fact, the extremely slow  decay of these
excitations indicate the difficulty of settling down the issue of the
existence of the internal mode. If simulations are not done very carefully
and runned for very long times, then 
this fact can lead to incorrect interpretations. 

As we have pointed out,  one can get oscillitory kink configurations by
constructing e\-xact solutions corresponding to the stationary
supperposition of a kink 
and one or many breathers. Such a case of a kink and a breather was named
`a wobble' by K\"alberman \cite{Kalberman}. In the case of the wobble the
frequency of oscillation cannot be greater than $1$, and the energy of
the oscillation goes to zero as the frequency approaches $1$. Boesch
and Willis \cite{Boesch} claimed to have seen an oscillatory mode of
the kink just above the phonon band, {\it i.e.} just above $1$. If that is
so, the wobble does not correspond to that mode. It is true however,
that the frequency of the wobble can go above $1$ by a Lorentz
boost. One has then to settle the issue of how precise the simulations
of Boesch and Willis were to separate this effect. If one considers
the exact stationary supperposition of a kink and two breathers one
can get frequencies of oscillations greater than $1$, as shown in
(\ref{twobreathers}). However, this is in a context different from
that discussed in the literature where the considered frequencies are
studied in 
the linear approximation. In addition, although our simulations have
shown that the kink plus breathers are stable solutions against the
discretisation, they can be pulled apart by scattering through a hole, as
mentioned at the end of section \ref{sec:wobblemain}. The fact that a kink
solution of the  equations of motion 
can involve many breathers makes the problem of the zero mode
difficult to resolve 
analytically (and in the literature 
it is discussed only in the `linear approximation') and almost
impossible to resolve numerically. 
The rich spectrum of the solutions and the appearance of many
breathers makes this task particularly 
hard to perform. It would be interesting to see whether these extra
breathers play a significant role 
in any physical applications of the model.

\appendix

\section{Appendix}

\subsection{Hirota's method}

In order to solve the sine-Gordon equation (\ref{two}) by the Hirota's
method we introduce the Hirota's $\tau$-function as
\be
\vp =  \frac{2\, i}{\beta}\; \log \frac{\tau_1}{\tau_0}.
\label{hirotatau}
\ee
Using light-cone coordinates
\be
x_{\pm}= \frac{1}{2}\( t \pm x\) \quad \qquad \pa_{\pm} =
 \pa_t\pm \pa_x \quad \qquad
\pa^2=\pa_t^2-\pa_x^2=\pa_{+}\pa_{-} 
\label{lightcone}
\ee
and replacing (\ref{hirotatau}) into (\ref{two}) we get
$$
\frac{\pa_{+}\pa_{-}\tau_0}{\tau_0} - \frac{\pa_{+}\tau_0\,
  \pa_{-}\tau_0}{\tau_0^2} -  
\frac{m^2}{4}\left[\(\frac{\tau_1}{\tau_0}\)^2-1\right] = 
\frac{\pa_{+}\pa_{-}\tau_1}{\tau_1} - \frac{\pa_{+}\tau_1\,
  \pa_{-}\tau_1}{\tau_1^2} -  
\frac{m^2}{4}\left[\(\frac{\tau_0}{\tau_1}\)^2-1\right] .
$$
Imposing both sides to vanish we get the Hirota's equations for the sine-Gordon
model:
\br
\tau_0 \, \pa_{+}\pa_{-}\tau_0 - \pa_{+}\tau_0\, \pa_{-}\tau_0 + 
\frac{m^2}{4}\(\tau_0^2-\tau_1^2\) &=&0, \nonu\\
\tau_1 \, \pa_{+}\pa_{-}\tau_1 - \pa_{+}\tau_1\, \pa_{-}\tau_1 + 
\frac{m^2}{4}\(\tau_1^2-\tau_0^2\) &=&0.
\label{hirotaeq}
\er
The solutions of (\ref{hirotaeq}) are obtained by the Hirota's ansatz  
\be
\tau_{\alpha} = 1+ \varepsilon\,\sum_{i=1}^N b_i^{(\alpha)} \, e^{\Gamma_i} + 
\varepsilon^2\,\sum_{i,j=1}^N  b_{ij}^{(\alpha)} e^{\Gamma_i+\Gamma_j} 
+ \varepsilon^3\,\sum_{i,j,k=1}^N b_{ijk}^{(\alpha)}
e^{\Gamma_i+\Gamma_j+\Gamma_k}\,+...\qquad \alpha =0,1,
\label{hirotaansatz}
\ee
where
\be
\Gamma_i = m \( z_i \, x_{+}-\frac{x_{-}}{z_i}\) 
\label{gammadef}
\ee
and where $z_i$ are arbitrary (complex) parameters. 
Replacing (\ref{hirotaansatz}) into (\ref{hirotaeq}) and expanding in
powers of $\varepsilon$, one obtains the coefficients $b^{(\alpha)}$'s
recursively. The series in (\ref{hirotaansatz})   truncates at
order $N$ leading to 
an exact solution, given by  
\br
\tau_{\alpha}&=& 1 + (-1)^{\alpha} \sum_{l=1}^N a_l \, e^{\Gamma\(z_l\)} + 
 \sum_{l_1<l_2=1}^N 
\(\frac{z_{l_1}-z_{l_2}}{z_{l_1}+z_{l_2}}\)^2\, 
 a_{l_1}\, a_{l_2} \, e^{\Gamma\(z_{l_1}\)+\Gamma\(z_{l_2}\)}+\ldots
\nonu\\
&+&(-1)^{\alpha}\sum_{l_1<l_2<l_3=1}^N 
\(\frac{z_{l_1}-z_{l_2}}{z_{l_1}+z_{l_2}}\)^2 
\(\frac{z_{l_1}-z_{l_3}}{z_{l_1}+z_{l_3}}\)^2
\(\frac{z_{l_2}-z_{l_3}}{z_{l_2}+z_{l_3}}\)^2 
a_{l_1} a_{l_2}  a_{l_3}
e^{\Gamma\(z_{l_1}\)+\Gamma\(z_{l_2}\)+\Gamma\(z_{l_3}\)} 
\nonu\\
&& \ldots + (-1)^{\alpha\,N} 
\prod_{k_1<k_2=1}^{N}
\(\frac{z_{k_1}-z_{k_2}}{z_{k_1}+z_{k_2}}\)^2 
\prod_{l=1}^N a_l \,e^{\Gamma\(z_{l}\)}
\qquad\qquad {\rm for}
\quad \alpha =0,1,
\label{nsolitonsolution}
\er
where $a_i$ are arbitrary parameters.

\subsection{The one-soliton solutions}
\label{sec:onesol}

The  one-soliton solutions correspond to $N=1$ and the following
choice of parameters in (\ref{nsolitonsolution})
\be
a_1 = i \, e^{-\varepsilon\,\gamma \, x_0} \qquad \qquad \qquad 
z_1 = \varepsilon\, e^{-\alpha} \qquad \quad \varepsilon = \pm 1\qquad
\alpha \; \; {\rm real}. 
\ee
Then the argument of the exponentials become
\be
\Gamma_1 = \varepsilon\, \gamma \( x-v\, t\) \qquad \qquad \gamma = m\,
\cosh \alpha =\frac{m}{\sqrt{1-v^2}}\qquad \qquad v =  \tanh \alpha. 
\ee
Therefore, (\ref{hirotatau}) and (\ref{nsolitonsolution}) gives 
\be
\vp = \frac{4}{\beta}\; {\rm ArcTan}\; \exp\left[\varepsilon\, \gamma \(
  x-v\, t-x_0\)\right]. 
\ee
\subsection{The breather solution}
\label{sec:breather}

The breather solution lies in the $2$-soliton sector, and it is
obtained by taking $N=2$, and the following choice of parameters in 
(\ref{nsolitonsolution})  
\be
z_1=e^{-\alpha+i\theta} \qquad \qquad z_2=z_1^* \qquad \qquad 
a_1=i\, \frac{e^{\eta+i\xi}}{\tan \theta} \qquad \qquad a_2=-a_1^*.
\ee
Then we have that $\Gamma_2=\Gamma_1^*$ and introducing ${\tilde
  \Gamma}_1=\Gamma_1+ \eta+i\xi$ we obtain
\be
{\tilde \Gamma}_1 = \Gamma_R +i\, \Gamma_I 
\ee
with
\be
\Gamma_R = \frac{m\, \cos \theta}{\sqrt{1-v^2}} \; \( x -
v\, t\) +\eta \qquad \qquad  
\Gamma_I = \frac{m\,\sin \theta}{\sqrt{1-v^2}} \(  t
-v\, x\) +\xi
\ee
and
\be
v= \tanh \alpha.
\ee
Thus
\be
\vp = \frac{4}{\beta}\; {\rm ArcTan}\, \frac{\({\rm cotan}\, \theta\)
  \, \cos \Gamma_I}{\cosh \Gamma_R }. 
\ee
If one now takes
$$
 \eta=v=0 \qquad \quad \xi = \frac{\pi}{2}
\qquad \quad \omega\equiv \sin \theta \qquad -\frac{\pi}{2}\leq \theta
\leq \frac{\pi}{2} 
$$ 
one gets 
\be
\vp = \frac{4}{\beta}\; {\rm ArcTan}\,\left[
  \frac{\sqrt{1-\omega^2}}{\omega}\;\frac{ \sin\( m\,\omega\, t\)
  }{\cosh\(m\, \sqrt{1-\omega^2}\; x\)}\right]. 
\ee

\subsection{The wobble or the kink with a breather}
\label{sec:wobbleapp}

To have a configuration describing a kink with a breather we take the
following values for the parameters of the $3$-soliton solution
($N=3$) in (\ref{nsolitonsolution})
\br
z_1&=&e^{-\alpha_B+i\theta} \qquad \qquad z_2=z_1^*
\qquad \qquad z_3 = \varepsilon e^{-\alpha_K}\nonu\\ 
a_1&=&i\, \frac{e^{\eta_B+i\xi_B}}{\tan \theta} \qquad \qquad 
a_2=-a_1^*  
 \qquad \qquad a_3 =  i\, e^{\eta_K}.
\er
Then one gets 
\be
{\tilde \Gamma_3} = \varepsilon\, \gamma_K \( x-v_K\, t\) + \eta_K
\qquad \qquad \gamma_K = m\, \cosh \alpha_K
\qquad \qquad v_K =  \tanh \alpha_K 
\ee
and again one has $\Gamma_2=\Gamma_1^*$. Then  introducing ${\tilde
  \Gamma}_1=\Gamma_1+ \eta_B+i\xi_B$, one obtains 
\be
{\tilde \Gamma}_1 = \Gamma_R +i\, \Gamma_I 
\ee
with
\br
\Gamma_R &=& \frac{m}{\sqrt{1-v_B^2}}\, \cos \theta \; \(
x - v_B\, t\)+ \eta_B \nonu\\  
\Gamma_I &=& \frac{m}{\sqrt{1-v_B^2}}\,\sin \theta \( 
t -v_B\, x\) +\xi_B\nonu
\er
and $v_B= \tanh \alpha_B$. Next we define
\be
\(\frac{z_1-z_3}{z_1+z_3}\)^2 = \rho\, e^{i\, \phi} 
\ee
with
\be
\rho =\frac{\cosh\(\alpha_B-\alpha_K\)-\varepsilon\, \cos
  \theta}{\cosh\(\alpha_B-\alpha_K\)+\varepsilon\, \cos \theta} 
\qquad \hbox{and}\qquad
\phi= 2\, {\rm ArcTan}\, \frac{\varepsilon\,\sin
  \theta}{\sinh\(\alpha_B-\alpha_K\)}. 
\ee
Then
\be
\vp = \frac{4}{\beta}\; {\rm ArcTan}\;\frac{\left[ 2\,\({\rm
      cotan}\theta\)\,\cos \Gamma_I +   
e^{{\tilde \Gamma}_3}\( e^{-\Gamma_R} + 
 \rho^2\; e^{\Gamma_R}\)\right]}{\left[ \(e^{-\Gamma_R} +   e^{\Gamma_R} \) 
- 2\,\({\rm cotan}\theta\)\,\rho\, e^{{\tilde \Gamma}_3} \,
\cos\(\Gamma_I+\phi\)\right]} .
\label{kink+breather}
\ee


If one now takes
$$
\eta_K=\eta_B=v_B=v_K=0 \qquad \quad
\xi_B=\frac{\pi}{2}  
$$
and denotes
$$
\omega\equiv \sin \theta \qquad -\frac{\pi}{2}\leq \theta \leq \frac{\pi}{2}
$$
then
\be
\rho =\frac{1-\varepsilon\, \sqrt{1-\omega^2}}{1+\varepsilon\,
  \sqrt{1-\omega^2}} 
\qquad \qquad
\phi= \pm \, \pi 
\ee
and so
\be
\vp = \frac{4}{\beta}\; {\rm ArcTan}\;\frac{\left[
    \frac{\sqrt{1-\omega^2}}{\omega}\,\sin \(m\, \omega\, t\)  +   
\frac{1}{2}\, e^{\varepsilon\, m\, x}\( e^{-m\,\sqrt{1-\omega^2}\, x} + 
 \rho^2\; e^{m\,\sqrt{1-\omega^2}\,x}\)\right]}{\left[  \cosh\(
    m\, \sqrt{1-\omega^2}\,x\) 
+ \frac{\sqrt{1-\omega^2}}{\omega}\,\rho\, e^{ \varepsilon\, m\, x} \,
\sin\(m\,\omega\, t\)\right]}. 
\ee

\subsection{The kink with two breathers}
\label{sec:kinktwobreathers}

To get a field configuration describing a kink and two breathers that
are all at rest and located at the same position we take the
parameters of the $5$-soliton solution ($N=5$) in (\ref{nsolitonsolution}) as
\be
z_1 = e^{i\,\theta_1} \qquad z_2 = z_1^* \qquad z_3 =
e^{i\,\theta_2} \qquad z_4 = z_3^* \qquad  
z_5 = \varepsilon
\ee
and
\be
a_1=-a_2= -{\rm cotan}\, \theta_1 \qquad \qquad 
a_3=-a_4= -{\rm cotan}\, \theta_2 \qquad \qquad
a_5= i.
\ee
Then setting $m=\beta=1$, one gets 
\br
\Gamma_1= \Gamma_2^*= x\; \cos \theta_1  + i\, t\; \sin \theta_1 \qquad
\Gamma_3= \Gamma_4^* =  x\; \cos \theta_2  + i\, t\; \sin \theta_2 \qquad
\Gamma_5= \varepsilon\; x.
\er
In addition, we have 
\br
\(\frac{z_1-z_2}{z_1+z_2}\)^2 &=& - \; \tan^2 \theta_1, \qquad 
\(\frac{z_3-z_4}{z_4+z_4}\)^2  =  - \;\tan^2 \theta_2,
\nonu\\
\(\frac{z_1-z_5}{z_1+z_5}\)^2 &=& \(\frac{z_2-z_5}{z_2+z_5}\)^2=
- \;\frac{1-\varepsilon\, \cos \theta_1}{1+\varepsilon\, \cos
  \theta_1} \equiv \rho_1, 
\nonu\\
\(\frac{z_3-z_5}{z_3+z_5}\)^2 &=& \(\frac{z_4-z_5}{z_4+z_5}\)^2=
-\; \frac{1-\varepsilon\, \cos \theta_2}{1+\varepsilon\, \cos
  \theta_2} \equiv \rho_2,
\nonu\\
\(\frac{z_1-z_3}{z_1+z_3}\)^2 &=& \(\frac{z_2-z_4}{z_2+z_4}\)^2=
- \;\frac{1- \cos \(\theta_1-\theta_2\)}{1+ \cos\(\theta_1-
  \theta_2\)}\equiv \sigma_{12}^{(-)}, 
\nonu\\
\(\frac{z_1-z_4}{z_1+z_4}\)^2 &=& \(\frac{z_2-z_3}{z_2+z_3}\)^2=
-\; \frac{1- \cos \(\theta_1+\theta_2\)}{1+ \cos\(\theta_1+
  \theta_2\)}\equiv \sigma_{12}^{(+)}. 
\er
The $\tau$-functions now become ($\alpha =0,1$)
\br
\tau_{\alpha}&=& 1
\nonu\\
&+& (-1)^{\alpha} \, i\;\left[-2\, {\rm cotan}\, \theta_1\, e^{x\;
    \cos \theta_1}\, \sin\(t\; \sin \theta_1\)  
 -2\, {\rm cotan}\, \theta_2\, e^{x\; \cos \theta_2}\, \sin\(t\; \sin
 \theta_2\)   
+ e^{\varepsilon\, x}\right]
\nonu\\
&+& e^{2\,x\; \cos \theta_1} + e^{2\,x\; \cos \theta_2}
+ 2\, \rho_1\, {\rm cotan}\, \theta_1\, e^{\varepsilon\, x}\, e^{x\;
  \cos \theta_1}\, \sin\(t\; \sin \theta_1\)  
\nonu\\
&+&  2\, \rho_2\, {\rm cotan}\, \theta_2\, e^{\varepsilon\, x}\,
e^{x\; \cos \theta_2}\, \sin\(t\; \sin \theta_2\)  
\nonu\\
&+& 2\, \sigma_{12}^{(-)}\, {\rm cotan}\, \theta_1\,{\rm cotan}\,
\theta_2\,e^{x\; \(\cos \theta_1+\cos \theta_2\)}\,  
\cos\(t\, \(\sin \theta_1+\sin \theta_2\)\)
\nonu\\
&-& 2\, \sigma_{12}^{(+)}\, {\rm cotan}\, \theta_1\,{\rm cotan}\,
\theta_2\,e^{x\; \(\cos \theta_1+\cos \theta_2\)}\,  
\cos\(t\, \(\sin \theta_1-\sin \theta_2\)\)
\nonu\\
&+&(-1)^{\alpha} \, i\;\left[-2\, {\rm cotan}\, \theta_2\;
  \sigma_{12}^{(+)}\, \sigma_{12}^{(-)}\, 
e^{x\; \(2\,\cos \theta_1+\cos \theta_2\)}\, \sin\(t\; \sin \theta_2\) 
+ \rho_1^2   \; e^{ 2\, x\; \cos \theta_1} \;  e^{\varepsilon\, x}
\right.\nonu\\ 
&-&\left.  2\, {\rm cotan}\, \theta_1\; \sigma_{12}^{(+)}\, \sigma_{12}^{(-)}\,
e^{x\; \(\cos \theta_1+2\,\cos \theta_2\)}\, \sin\(t\; \sin \theta_1\)   
+ \rho_2^2   \; e^{ 2\, x\; \cos \theta_2} \;  e^{\varepsilon\, x}
\right. \nonu\\ 
&+& \left. 2\, \sigma_{12}^{(-)}\, \rho_1\, \rho_2\, {\rm cotan}\,
\theta_1\,{\rm cotan}\, \theta_2 
\, e^{\varepsilon\; x}\, e^{x\; \(\cos \theta_1+\cos \theta_2\)}\, 
\cos\(t\, \(\sin \theta_1+\sin \theta_2\)\)\right. \nonu\\
&-& \left. 2\, \sigma_{12}^{(+)}\, \rho_1\, \rho_2\, {\rm cotan}\,
\theta_1\,{\rm cotan}\, \theta_2 
\, e^{\varepsilon\; x}\, e^{x\; \(\cos \theta_1+\cos \theta_2\)}\, 
\cos\(t\, \(\sin \theta_1-\sin \theta_2\)\)
\right]\nonu\\
&+& \(\sigma_{12}^{(+)}\,\sigma_{12}^{(-)}\)^2 \, e^{2\;x\; \(\cos
  \theta_1+\cos \theta_2\)} \nonu\\ 
&+&  2\, \sigma_{12}^{(+)}\,\sigma_{12}^{(-)}\, \rho_1^2\, \rho_2\,
    {\rm cotan}\, \theta_2\, e^{\varepsilon\, x}\,  
e^{x\; \(2 \cos \theta_1+\cos \theta_2\)}\, \sin\(t\; \sin \theta_2\) \nonu\\
&+&  2\, \sigma_{12}^{(+)}\,\sigma_{12}^{(-)}\, \rho_1\, \rho_2^2\,
{\rm cotan}\, \theta_1\, e^{\varepsilon\, x}\,  
e^{x\; \( \cos \theta_1+2\,\cos \theta_2\)}\, \sin\(t\; \sin \theta_1\) 
\nonu\\ 
&+&(-1)^{\alpha} \, i\;
\(\sigma_{12}^{(+)}\,\sigma_{12}^{(-)}\,\rho_1\, \rho_2\)^2\;
e^{\varepsilon\, x}\,  
e^{2\,x\; \( \cos \theta_1+\cos \theta_2\)}.
\er

This gives us the expression mentioned in section \ref{sec:kink+twobreathers}.



\vspace{2 cm}


{\bf Acknowledgements:}  This work was performed when WJZ visited the
University of S\~ao Paulo in S\~ao Carlos. 
His visit was supported by a grant from FAPESP which
 is gratefully acknowledged.
WJZ also wishes to thank the University of S\~ao Paulo in S\~ao Carlos for
its hospitality. We both thank Prof. F. C. Alcaraz for the use of the
computer facilities. WJZ also thanks  
Gilberto Nakamura for the help with computers in S\~ao Carlos and for
making him addictive to the Brazilian coffee.

\end{document}